# Tunable photon-induced spatial modulation of free electrons

Shai Tsesses[1,‡], Raphael Dahan[1,‡], Kangpeng Wang[1], Ori Reinhardt[1], Guy Bartal[1] and Ido Kaminer[1]

[1]Andrew and Erna Viterbi Department of Electrical Engineering, Technion, Israel Institute of Technology, 32000 Haifa, Israel

[‡]These authors contributed equally

**Spatial modulation of electron beams is an essential tool for various applications such as nanolithography and imaging[1–3], yet its implementations[4–11] are severely limited and inherently non-tunable. Conversely, light-driven electron spatial modulation[12–15] could potentially allow arbitrary electron wavefront shaping[16–18] via the underlying mechanism of photon-induced near-field electron microscopy (PINEM)[19–21]. Here, we present tunable photon-induced spatial modulation of electrons through their externally-controlled interaction with surface plasmon polaritons (SPPs). Using recently developed methods of shaping SPP patterns[22,23], we demonstrate a dynamic control of the electron beam with a variety of high-quality electron distributions. Intriguingly, by utilizing the intrinsic interaction nonlinearity[24], we attain the first observation of 2D spatial Rabi oscillations[25,26] and generate electron features below the SPP wavelength. Our work paves the way to on-demand electron wavefront shaping at ultrafast timescales, with prospects for aberration correction, nano-fabrication and material characterization.**

Current technologies for spatial modulation of coherent electron beams are based on three enabling technologies: tailored electro- and magneto-static fields[4,5], applying force to change electron trajectories; apertures[6] or binary holograms[7], nullifying electron transmission at pre-selected locations; and electron phase plates[8] or holograms[9,10], utilizing variations in thin-film thickness to shape the phase-front of the electron beam. Nevertheless, most contemporary methods lack the means to actively change the electron beam shape, and even state-of-the-art electron beam shaping technologies[11,27,28] are still quite restricted in their tunability, operation speed and scalability. Since tunable spatial modulation of beams is a capability that already revolutionized the fields of optics[29] and acoustics[30], breakthroughs advancing this enabling technology in electron optics are expected to facilitate potential discoveries and promote future applications.

A promising new approach to spatially modulate electrons relies on an entirely different physical mechanism – the interaction of free electrons with light, either in the presence of nanophotonic platforms[13–15] or through the ponderomotive force in free-space[12]. Such an interaction produced basic electron patterns in the image or diffraction plane of an electron microscope[12–15], inspiring several theoretical proposals[16–18] for the design of a nearly-arbitrary, light-driven electron spatial modulators. This concept, however, has never been demonstrated experimentally.

Our work takes the next step in this direction, displaying complex electron patterns that can be controlled by tuning external light properties and by using the inherently nonlinear nature of the free-electron–light interaction. We generate complex, light-driven electron distributions, such as various Bessel beams and a vortex array, via a patterned gold-coated silicon nitride membrane placed inside an ultrafast transmission electron microscope (UTEM). We continuously tune the shape of an electron in a superposition of Bessel modes, controlling their relative weights and showcasing the potential to adjust the orbital angular momentum of electrons in real-time. The nonlinear nature of free-electron–light interactions unlocks a novel type of electron spatial modulation: 2D spatial Rabi oscillations[25,26], obtained by adjusting the incident laser intensity and pulse width. Our findings are not only capable of impacting state-of-the-art electron imaging and characterization techniques, but are an important step toward fully-programmable electron beams, whose applications range from adaptive aberration correction in electron microscopes to on-demand generation of masks for electron beam lithography.

Our scheme is based on the interaction of fast, paraxial electrons with ultrafast SPP fields, which can be described by the following equation[20,21,31]:

$$\left[U_0 - \hbar v\left(i\frac{\partial}{\partial z} + k_0\right) - \frac{ie\hbar}{\omega}\left(e^{-i\omega t}E_z(x,y,z,t) - e^{i\omega t}E_z^*(x,y,z,t)\right)\right]\psi = i\hbar\frac{\partial \psi}{\partial t} \quad (1)$$

where $v$ is the electron velocity; $k_0$ is the electron initial wave-vector; $U_0$ is the electron initial energy; $E_z, E_z^*$ are the SPP electric field phasor in the direction of electron propagation and its complex conjugate; $\omega$ is the central frequency of the SPP field; $e$ is the electron charge; $\hbar$ is the reduced Plank constant; and $\psi$ is the electron wavefunction. Eq. 1 shows that the electron wavefunction is directly influenced by the shape of the electric field $E_z$.

Assuming that the SPP field has a slowly varying temporal envelope[26], the electric field acts on the electron wavefunction in a similar manner to a phase mask[24,31]:

$$\psi = \psi_0 \sum_{l=-\infty}^{\infty} J_l\left(2|g(x,y)|\right) e^{il\arg(-g(x,y))} e^{il\omega\left(\frac{z}{v}-t\right)} = \psi_0 e^{2i|g(x,y)|\sin\left(\omega\left(t-\frac{z}{v}\right)-\arg(g(x,y))\right)} \quad (2)$$

where $\psi_0$ is the initial electron wavefunction; and $g(x,y) = \frac{e}{\hbar\omega}\int_{-\infty}^{\infty} E_z(x,y,z)e^{-i\frac{\omega}{v}z}dz$ is the free-electron–light interaction strength, containing the transverse $(x,y)$ spatial dependence of the modulation. Eq. 2 indicates that the probability of the electron in point $(x,y)$ to absorb or emit $l$ photons is $P_l(x,y) \propto J_l^2\left(2|g(x,y)|\right)$. The index $l$ thus denotes the interaction order, and by post-selecting electrons of a specific energy range, one can isolate a specific interaction order or several of them. Then, regardless of the interaction strength, the modulation is no longer confined to the electron phase but involves its amplitude as well. An illustration of this amplitude modulation concept is given in fig. 1.

As fig. 1 explains, the wavefunction dependence on the parameter $g$ defines two regimes of free-electron–light interaction: an approximately linear interaction and an inherently nonlinear one. In the weak interaction limit ($g \ll 1$), the free-electron–light interaction can obviously be linearized. Alternatively, for laser pulses shorter than the temporal extent of the electron wavefunction, the same electron experiences a range of values for the interaction strength, resulting in a mixture of different interaction orders. In both cases, the electron probability distribution after interaction shows a similar dependence on the field intensity, roughly scaling as[20,21] $P(x,y) \propto |g(x,y)|^2 \propto |E_z(x,y)|^2$. In this linear interaction regime, which was used in most PINEM experiments thus far[13,14,19,26,32–35], the spatial modulation of the electron is directly derived from the spatial shape of the electro-magnetic near-field, such as the one obtained by surface-plasmon polaritons (SPPs).

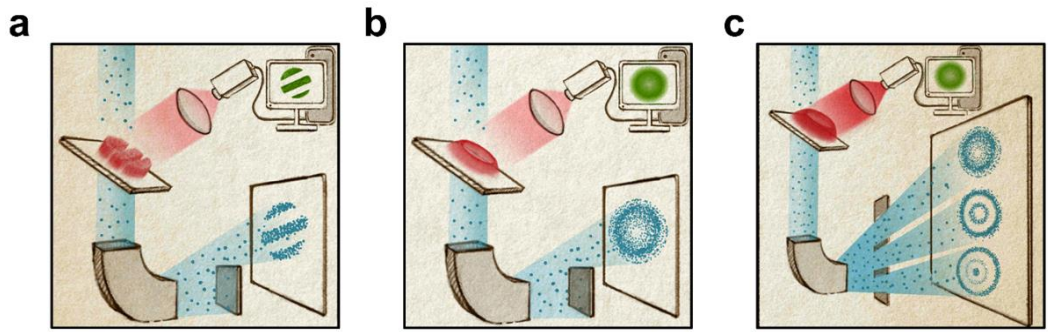

**Figure 1 | Tunable photo-induced free electron spatial amplitude modulation: concept illustration. (a)** In a similar fashion to photon-induced near-field electron microscopy experiments[19,26,32,33,36], a laser excitation is used to create a specific near-field distribution. In an ideal scenario, the required parameters of the laser illumination for a certain near-field shape are controlled automatically, via a computer software. An electron beam impinges the sample along with the laser excitation and is affected by the near-field. After propagating through various electron optical components (not shown in the illustration), the electron beam enters an electron energy loss spectrometer, whose output energies are post-selected in advance to collect electrons that gained energy through their interaction with the near-field. Those are then imaged onto a camera after the spectrometer, revealing that the electrons chosen now possess the near-field shape. **(b)** In the linear free-electron–light interaction regime, one can shape the electron beam directly by controlling the electromagnetic near-field, generating arbitrary electron shapes. **(c)** In the nonlinear free-electron–light interaction regime, defined by a larger light intensity, one can generate different electron beam shapes by post- selecting specific electron energies, even without changing the near-field shape, as the spatial information embedded in each electron energy is not the same. Thus, both the shape of the near-field (in the linear regime) and its intensity (in the nonlinear regime) are tunable degrees-of-freedom with which to spatially modulate electron beams. Drawings were produced by SimplySci.

Before delving into the nonlinear interaction regime, we will first present our demonstration of spatial electron modulation in the linear one. To this end, we shape long-range SPPs in a thin gold-coated silicon nitride membrane using specially designed coupling slits[37–39] that enable dynamic control over the electron wavefunction by altering the SPP field through the incident illumination. The various experimentally tunable parameters for this setting are given in fig. 2b.

The SPP field patterns are generated by weakly focused femtosecond laser pulses that impinges the coupling slits inside a UTEM (fig. 2a; see extended data fig. 1 and the Methods section for further details about the sample, its electromagnetic properties and the experimental system). Femtosecond free electron pulses arrive at the sample simultaneously with the laser pulses and pass through it. The electrons impinge an excitation slit carved into the membrane in its central uniform area, such that the entire electron shaping is produced by the SPP field and not by the aperture created by the slit. The electrons arrive simultaneously with the laser pulse. We energy-filter the post-interaction electrons with a wide window at the gain side of the electron energy spectrum, effectively collecting the signal from all electrons that gained energy ($l > 0$).

The results are summarized in fig. 2c-e, presenting various examples of electron spatial modulation enabled by shaping the SPPs: a 1$^{st}$-order Bessel electron vortex (fig. 2c); a hexagonal electron vortex array (fig. 2d); and a hexagonal foci array (fig. 2e). Each measurement is accompanied by the calculated electron probability distribution after interaction, showing a good match between experiment and theory (full calculation details, as well as the procedure for processing the raw electron microscope images are given in the Methods section). We note that our photon-induced spatial modulation matches in quality to the state-of-the-art passive electron spatial modulation schemes[40,41], while going beyond the single, localized vortex previously generated through free-electron–light interactions[14]. It is further important to note that the hexagonal electron foci lattice is generated by a lattice of SPP skyrmions in the electric field[42,43], whose influence on the amplitude of an incident electron beam is surprisingly similar to their magnetic counterparts' influence on the beam's phase[44].

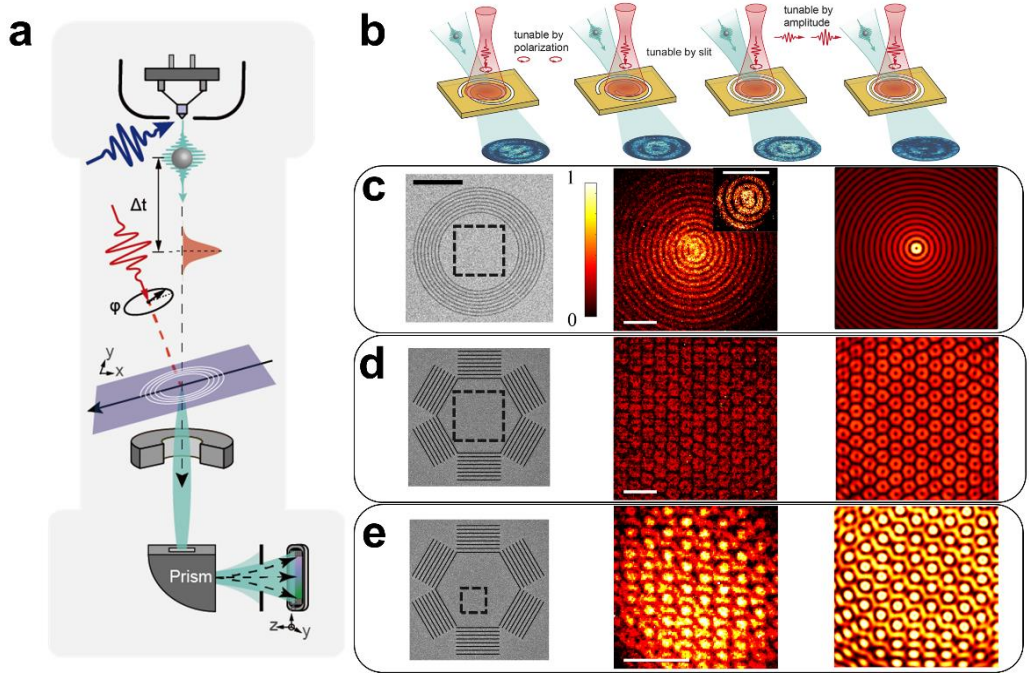

**Figure 2 | Photon-induced spatial modulation of free electrons – measurements and experimental setup. (a)** An illustration of the ultrafast transmission electron microscope (see the Methods section for details). After interaction with the generated near-field in the sample, the electrons are imaged at the output plane of the membrane. The electron signal is post-selected using energy-filtered imaging to show the electron distribution of electrons gaining specific energies (or their sum). **(b)** Various degrees of tunability for spatial modulation of electron wavefunctions in our experiment. In the linear interaction regime, the near-field SPP amplitude within a designed coupling slit is imprinted onto the post-interaction electrons. Changing the polarization of the incident laser pulse modifies the shape of the guided mode's electric field in-turn, facilitating active control over the electron spatial distribution. Varying the shape of the coupling slit also changes the electron shape by the change in the SPP field, yet this degree of freedom is fixed for each sample. Increasing the laser field amplitude while extending its duration induces nonlinearity in the electron light-interaction, causing electrons that gained a different amount of energy obtain different shapes and adding another degree of control. **(c)-(e)** Measured electron probability distributions after interaction with the near field of various plasmonic patterns generated by coupling slits with different geometries, including **(c)** a $1^{st}$-order Bessel electron vortex, **(d)** a hexagonal electron vortex array and **(e)** a hexagonal electron foci array. Each electron distribution is accompanied by a scanning electron microscope (SEM) micrograph of the coupling slit used to generate it. The measurement area is marked in every micrograph by a dashed square. Inset **(c)** is an image of the measured distribution, taken at a larger magnification, where the vortex singularity is more clearly visible. The measurements in **(c)-(e)** are complemented by their corresponding calculated probability distributions, which show a good match between theory and experiments and follow the shape of the plasmonic near-field[22,23,45] as expected in the linear interaction regime. The black scale bar (relevant to all SEM micrographs) corresponds to 10 microns. All white scale bars correspond to 2 microns.

After establishing the variety and quality of electron spatial modulation possible using our scheme, we demonstrate its tunability via external control of the near-field profile (fig. 3). Utilizing a spiral plasmonic coupling slit, we alter the boundary conditions of the SPP field as a function of the polarization of the incident laser pulse, which then determines the interference pattern within the slit. A gradual variation in the laser polarization results in the excitation of either one of two specific Bessel modes[38] or their superposition[22], which is imprinted upon the electron (see also supplementary movies 1 & 2). Slight distortions in the shape of the modes arise from the nonideal nature of the spiral slit excitation and the impure circular polarization in our system.

While the results in fig. 2,3 are achieved in the approximately linear interaction regime, our electron spatial modulation method is inherently nonlinear, opening new degrees of freedom for more complex spatial modulation schemes. The nonlinearity originates from two main mechanisms: 1. the electron absorbs and emits several SPP quanta in a stimulated matter and 2. the action of post-selecting specific electron energies (marked as the interaction order $l$ in Eq. 2).

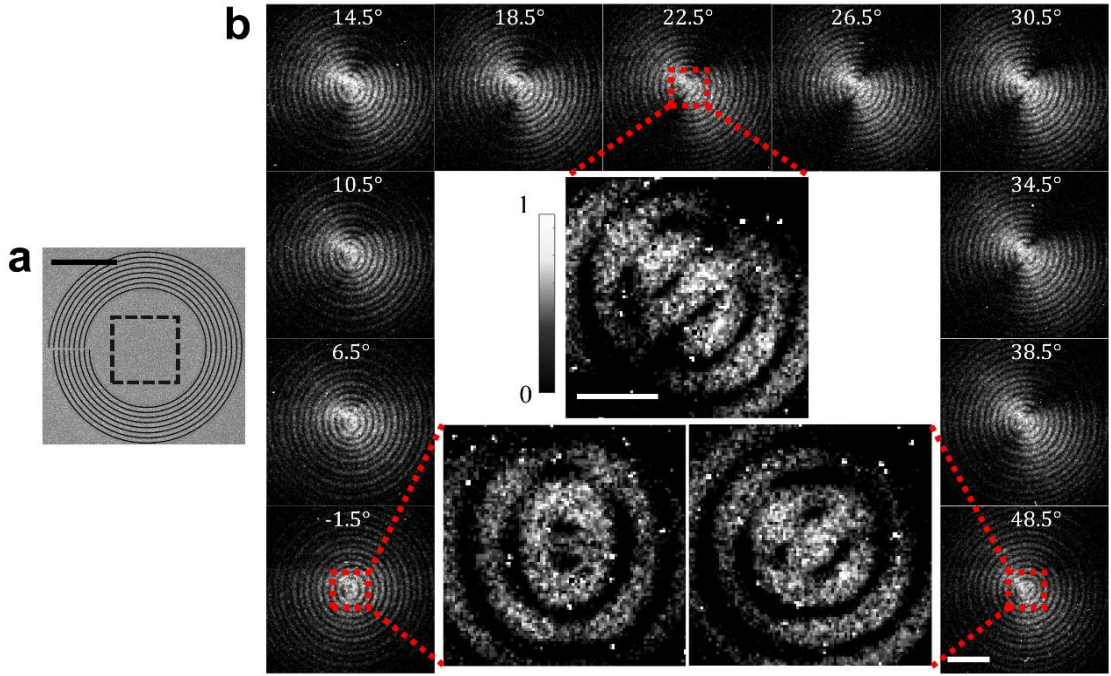

**Figure 3 | Spatial modulation of free electrons by active control of SPP boundary conditions. (a)** An SEM micrograph of the spiral coupling slit used for the measurements, with the measurement area marked by a dashed square. Inset scale bar is 10 microns. **(b)** Measured normalized probability distribution of post-interaction electrons as a function of the rotation angle of a $\lambda/4$ plate placed outside the UTEM (written above every panel), which controls the laser pulse polarization. As the $\lambda/4$ plate rotates, the electron gradually transforms form a second-order Bessel mode to a zero-order Bessel mode, existing in a superposition of both modes at any intermediate step. Three representative distributions are also presented in a larger magnification (with the measurement area marked by a dashed red line), showcasing the difference in shape. The short and long white scale bars correspond to 2.5 and 0.5 microns, respectively.

We investigate the properties of electron spatial modulation in the nonlinear interaction regime, showing that electrons with different energies possess different probability distributions. We utilize a circular coupling slit (as in fig. 2c) and attain the nonlinear regime by increasing both the intensity and temporal duration of the incident laser pulses. Figs. 4a,b present the measured probability distribution in each positive interaction order (at different magnifications), while fig. 4c shows a radial average comparison of all distributions. The key to visualize the different interaction orders is using narrow-band energy filtering around each of their energies ($l\hbar\omega$, see the Methods section and extended data fig. 2 for more details).

The electron probability distribution for each order $l$ in the nonlinear regime depends on the local amplitude of the near-field: Away from the slit's center, where the near-field amplitude is low, the probability distribution is monotonic with the interaction order, as in the linear regime. Toward the center of the slit, as the near-field amplitude increases, the ratio between the probability distributions oscillates as a function of the local near-field intensity. This phenomenon is known as Rabi oscillations in the coherent interaction of electrons and optical excitations, and was previously demonstrated in zero[25] and one[26,36] spatial dimensions. Our measurements record this phenomenon in 2D for the first time and with a wide transverse electron wavefunction (past demonstrations scanned a strongly-focused electron over a near-field pattern). Thus, unlike previous studies, our observed Rabi oscillations can be utilized to shape the electron wavefunction, as electrons gaining a different amount of energy possess a different spatial distribution.

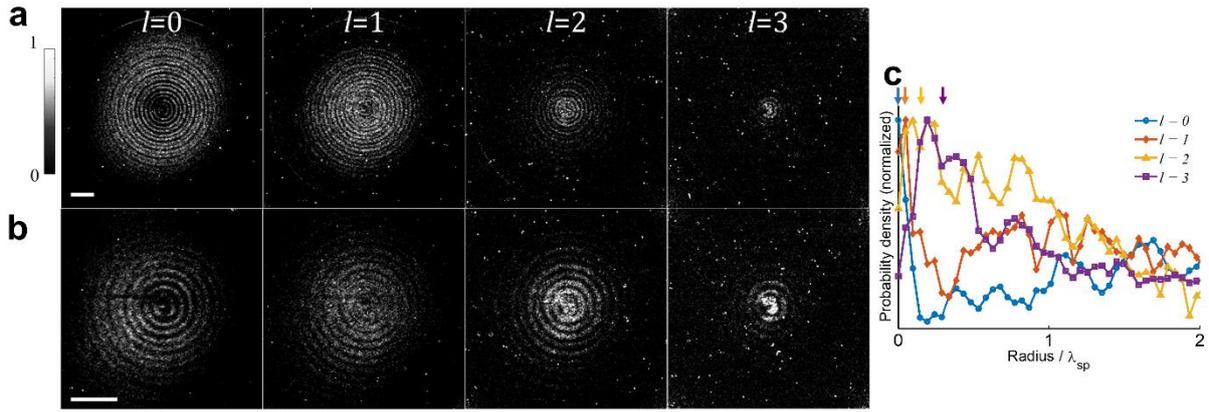

**Figure 4 | Intensity-tunable photo-induced spatial modulation of electrons in the nonlinear interaction regime. (a),(b)** Measured energy-filtered electron probability distribution for individual interaction orders $l$, defined in Eq. 2. Each electron distribution is presented both in a larger field of view **(a)** and at a larger magnification, exposing more delicate details **(b)**. In the larger field of view, it is evident that far away from the center of the pattern, the probability is monotonous with the interaction order (as in the linear regime). In the larger magnification images, close to the center of the pattern, the various probability distributions are inherently different and obtain maximal values at different locations, thus exhibiting 2D spatial Rabi oscillations between different electron energy levels. This phenomenon is more clearly visible by comparing a radial average of the normalized probability density at each order **(c)**, revealing that the full theory for the nonlinear free-electron–light interaction is necessary in this case. It is noticeable in both **(b)** and **(c)** that electrons undergoing no interaction ($l=0$) exhibit a central spot much smaller than the plasmonic wavelength ($\lambda_{sp}$), whose asymmetry is due to nonideal excitation of the near-field. One can also discern from **(a)** and **(b)** how the increasing interaction order gradually resembles the SPP field intensity, like the result in the linear regime (fig. 2c). Color-coded arrows in **(c)** mark the location of the main lobe in each interaction order, shifting to a larger radius as $l$ increases, until coinciding with the expected location of the near-field maximum for $l=3$. The short and long white scale bars in **(a)** and **(b)**, respectively, both correspond to 2 microns.

Interestingly, while the SPP wavelength limits the absolute *number* of controllable electron features regardless of the interaction regime, it limits the absolute *feature size* only in the linear interaction. In the nonlinear interaction regime, there is, in essence, no such limit. A prominent example of a subwavelength feature is apparent in the interaction order $l=0$ in fig. 4 (i.e., electrons gaining no energy in the interaction). The resulting probability distribution exhibits an elliptic spot at its center, whose peak-to-dip radius is up to 3 times smaller than the smallest possible spot in a 2D plasmonic near-field (varies between a fifth and a seventh of the SPP wavelength $\lambda_{sp}$). This feature is generated by the phase singularity at the center of the Bessel vortex near-field, which allows electrons to pass without interaction in an area whose size is determined by the near-field amplitude. Such a saturation phenomenon bears great similarity to the stimulated emission depletion (STED) mechanism, currently used in super-resolution fluorescence microscopy[46].

In conclusion, we showed that both passive and active spatial electron modulation are achievable through interactions of free-electron pulses with ultrafast surface plasmon polariton excitations in a metal-dielectric membrane. Intuitively, the goal of tunable spatial electron modulation is considered by many as analogous to spatial light modulators (SLM) in optics, shaping electrons instead of light. Interestingly, the mechanism our spatial modulation scheme is actually much more similar to acousto-optical deflection[47] with many degrees of freedom, which is commonly used for shaping beams of light[48]. Our approach promises more degrees of freedom for active spatial modulation, simpler fabrication than other current methods and can be performed at ultrafast timescales, allowing complete spatiotemporal modulation of electron wavefunctions[18]. Furthermore, our scheme is freely available to use in any ultrafast electron microscope or in other electron microscope variants[49], as it only requires standard fabrication processes, optical components and electron optics.

It is important to note that our scheme is readily capable of spatially modulating the *phase* of free-electrons along with their *amplitude*, utilizing the same plasmonic components (as shown in extended data fig. 4). Meaningful electron spatial phase modulation requires an electron transverse coherence longer than the SPP wavelength, which is quite possible with similar experimental parameters[14,15,32], though we did not directly observe it. Intriguingly, our results imply that with the addition of spatial phase modulation in the nonlinear interaction regime, the electron distribution may indeed contain a different angular momentum quantity in different areas, as predicted in [24].

Considering all the above, the electron shaping platform presented herein has the potential to immediately impact state-of-the-art electron imaging and characterization techniques. For example, a continuous and ultrafast tuning of an electron beam's angular momentum could enable characterization of chiral and magnetic materials[40], with high temporal and spatial resolutions. We also envision the generation of tunable nanometric apertures, smaller than those currently used in electron microscopes, in which electrons can pass freely only in certain areas defined by a pre-designed laser-intensity pattern (as in fig. 4). Furthermore, by matching an electron beam pattern to a desired lattice shape or symmetry (as in fig. 2d,e), electrons can be made to pass through the crystal with reduced scattering or excite only specific atoms in heterogeneous crystals.

Looking forward, our findings are a step toward fully-programmable electron beams with controllable wavefronts. Further improvements can be made by using a spatial light modulator[50] to shape the incoming laser pulses; employing multiple confined modes (as in a multimode membrane or at the interface of a mirror); or utilizing multi-frequency illumination[31]. In the end, complete automation of the photon-induced spatial modulation process promises real-time optimization of electron beam shapes, for continuous aberration correction of electron beams; on-demand generation of patterns for electron beam lithography; increased electron beam propagation inside disordered materials; and electron beam focusing inside samples beyond the mean free path restriction.

**Methods**

**Sample preparation.** We coated a 40 nm layer of Au atop a 30 nm $Si_3N_4$ TEM membrane (Norcada, NTX025) using sputter deposition (AJA International Inc., ATC 2200), with a 3 nm Ti adhesion layer. We use only the long-range surface plasmon polariton guided wave in our setup, which is localized at the sample boundaries and is only weakly affected by the adhesion layer (see extended data fig. 1a for the sample illustration and mode shape). After fabrication of the Au-coated membrane, plasmonic coupling slits are milled into it using a focused ion beam (FEI, Helios NanoLab DualBeam G3 UC), etching only through the Au coating, leaving the $Si_3N_4$ layer intact. The plasmonic coupling slits were optimized for broadband operation around an excitation wavelength of 730 nm. A top view of the entire sample from the Au side is given in extended data fig. 1b.

**Experimental setup.** We performed the measurements using the experimental setup described in [26]. Laser pulses are frequency-quadrupled to induce photoemission of electron pulses from the tip source of the electron microscope. These electrons are then accelerated and impinge a nanostructured membrane in sync with laser pulses that generate a plasmonic near-field interacting with the electrons. Laser pulses impinging the sample were tuned to a central wavelength of 730 nm by an optical parametric amplifier (Light Conversion, Orpheus), and their polarization is externally tunable via a broadband λ/4 plate. For the experiments described in fig. 2,3, we utilized laser pulses of ~220 fs (FWHM), at 1 MHz repetition rate and with an energy of ~25 nJ (25 mW average power), while Electron energy filtered imaging (EFTEM) was performed with a ~10 eV wide slit at the gain side of the energy loss spectrometer, practically observing all electrons that gained energy in their interaction. For the experiment described in fig. 4, the laser pulses were stretched using N-BK7 rods (Edmund Optics, Light Pipe) to ~620 fs (FWHM) and their energy was increased to ~50 nJ (50 mW average power). EFTEM was performed with a ~1eV window, centered around multiples of the central laser pulse frequency (0eV, -1.7eV, -3.4 eV and -5.1 eV), such that only electrons at a certain interaction order were imaged (with a small residual signal from other interaction orders). Electron pulse width and energy width were ~350 fs and ~1 eV for all experiments. The electron energy filtering scheme in each part of the experiment is illustrated in extended data fig. 2. The sample was tilted at an angle of ~4.5° to allow normal incidence of the laser pulses. Changes as small as 0.1° to this angle create aberrations in the shape of the spatially modulated electrons (see supplementary movie 3, where the sample tilt is varied between 4.1° and 4.9°), which provides another degree of freedom for controlling the electron wavefunction.

**Theory of free-electron–light interaction.** Our theoretical formalism is based on the one derived in [31]. We describe the interaction of a free electron and an electromagnetic wave by using the time-dependent Schrödinger equation:

$$\left[ \frac{(\vec{p}+e\vec{A})^2}{m} - eV \right]\psi = i\hbar \frac{\partial \psi}{\partial t}$$

where $\vec{p}$ is the momentum operator of the electron; $e$ is the elementary charge; $m$ is the electron mass; $\vec{A}$ is the electromagnetic vector potential; $V$ is the electromagnetic scalar potential; and $\psi$ is the electron wavefunction.

We apply the same assumptions in solving the above-mentioned equation as in all other papers on ultrafast transmission electron microscopy[20,21]. In short, the electron is highly paraxial, such that its longitudinal and transverse dynamics are decoupled, and the entire transverse dynamics can be neglected. This allows us to introduce the relativistic correction to the equation by simply replacing $m \to \gamma m$ (where $\gamma$ is the relativistic Lorentz factor). As a consequent of these conventional conditions, we can reach the form given in Eq. 1.

To solve Eq. 1, it is customary[24] to assume that the electron is made up of a carrier wave and a slowly varying envelope, which can be decomposed to harmonics of the frequency of the electromagnetic wave:

$$\psi = e^{i\left(k_0 z - \frac{U_0}{\hbar} t\right)} \phi_0\left(\vec{r} - vt\hat{z}\right) \sum_l e^{il\omega\left(\frac{z}{v} - t\right)} f_l\left(\vec{r}\right)$$

Where $U_0, k_0$ are the initial energy and momentum of the paraxial electron, respectively; $\hbar$ is the reduced plank constant; $\phi_0$ is a slowly-varying envelope function; $\omega$ is the frequency of the electromagnetic wave; $v$ is the initial electron velocity; $\vec{r} = (x, y, z)$ is the coordinate vector, with $z$ being the longitudinal coordinate; and $l, f_l$ are the interaction order and its spatially dependent coefficient, with negative orders denoting emission and positive orders denoting absorption.

When solving Eq. 1 with the above assumption for the wavefunction, one can extract an equation for the spatially dependent coefficient of each interaction order:

$$\frac{\partial f_l(\vec{r})}{\partial z} = \frac{e}{\hbar\omega}\left(E_z^* e^{i\omega\frac{z}{v}} f_{l+1}(\vec{r}) - E_z e^{-i\omega\frac{z}{v}} f_{l-1}(\vec{r})\right)$$

Which results in the final expression:

$$f_l(\vec{r}) = e^{il\arg(-g(x,y))} J_l\left(2|g(x,y)|\right)$$

where $g(x, y) = \frac{e}{\hbar\omega} \int_{-\infty}^{\infty} E_z(x, y, z) e^{-i\frac{\omega}{v} z} dz$ is the free-electron–light interaction strength; $J_l$ is the $l$-th order Bessel function of the first kind; and $E_z$ is the electric field phasor in the direction of electron propagation (the out-of-plane direction of our sample). The longitudinal dependence of the SPP field in our system is given in extended data fig. 1a, and a representative transverse distribution is given in extended data fig. 3a,b.

Under the approximation that $g \ll 1$, the coefficients of all interaction orders are negligible except for the first, and thus the transverse distribution of electrons gaining energy in the interaction is $P_T \approx |f_1(x, y)|^2$, $f_1(x, y) = 2|g(x, y)|e^{i\arg(-g(x,y))}$. Consequently, the transverse electron probability distribution follows the spatially-dependent intensity of the SPP field (see extended data fig. 3c,d). A similar scaling also appears for larger values of $g$, provided that the electron pulse duration is longer than the laser pulse duration. Both of these cases form the regime we refer to as "the linear interaction regime".

It is important to note that the spatial electron modulation in our work *does not need* to assume any transverse coherence for the interacting electrons. Transverse coherence is needed for spatial modulation of the electron in the *diffraction plane*. Our work does not include measurements in the diffraction plane and thus does not provide any experimental evidence for electron transverse coherence. Previous experiments demonstrated PINEM interactions on electrons with satisfactory transverse coherence to show how light can induce electron orbital angular momentum[14]. Therefore, the current concepts of spatial electron modulation should directly apply to electrons with wider transverse coherence in similar experimental setups. The transverse coherence can generally be improved by increasing the size of the electron spot (at the cost of lower signal)[14,15,32].

**Image acquisition and processing.** Images of the electron spatial distribution were acquired by a camera mounted on an electron energy loss spectrometer (Gatan). The images in fig. 2,3 were acquired using an exposure time of 120 s per image, while the exposure time in fig. 4 was 180 s per image. Two main sources of noise exist in every image: background noise and random flaring of detector pixels due to the detection mechanism (scintillation measured through a CCD detector). As a result, the raw detector images in our experiment are quite noisy, and the detector software usually performs substantial automated image processing on the raw image data. We exported the raw data (extended data fig. 4a) and performed the image processing ourselves.

Correcting the presented image for random pixel flaring requires contrast manipulation, which is performed by considering the distribution of the image histogram and choosing lower and upper bounds (extended data fig. 4b). We chose the average pixel signal as the lower bound, while the upper bound was chosen to be the highest pixel value 5 standard deviations from the average (this number changes for each image). Background noise may be removed by equalization – reducing the average pixel value from each pixel. A representative result of the full image processing is given in extended data fig. 4c, matched against the image produced by the detector software (extended data fig. 4d).

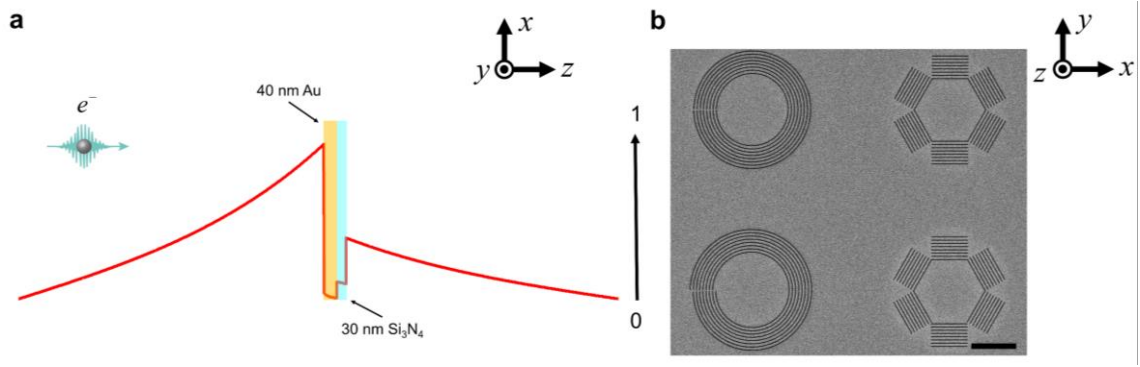

**Extended Data Figure 1 | Sample design and fabrication. (a)** Illustration of the sample used in the experiments, overlaid with the cross-section of the long-range surface plasmon polariton, showing its electric field amplitude in the direction of electron propagation ($|E_z|$). The direction from which the electron impinges the sample also appears in the illustration, as well as an axis for the amplitude strength. **(b)** A SEM micrograph of the various plasmonic coupling slits used in our experiments, which were optimized for broadband operation around an excitation wavelength of 730 nm and milled into the gold layer of the sample (scale bar is 10 microns). The coordinate system of the experiment appears inset **(a)** and **(b)**, rotated to fit the observation direction.

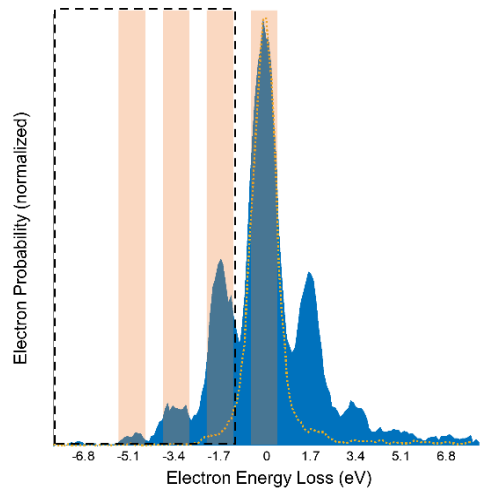

**Extended Data Figure 2 | Electron energy filtering schemes used in the experiment.** The figure shows a representative measurement of the electron energy loss spectrum (EELS) measured in our experiment (blue area), with visible peaks at integer multiples of the laser pulse (~1.7 eV). The measured EELS of the electron without laser pulse excitation is given by the dotted gold curve. For the measurements performed in figs. 2-3, we filter electrons that gained energy, as marked by the dashed black frame, effectively adding up all positive free-electron–light interaction orders. For the measurements performed in fig. 4, we filter electrons that underwent interactions of specific orders, as marked by the light orange rectangles (each with a ~1eV energy width). The energy resolution of our measurement was 0.1 eV, set by the configuration of the electron energy loss spectrometer.

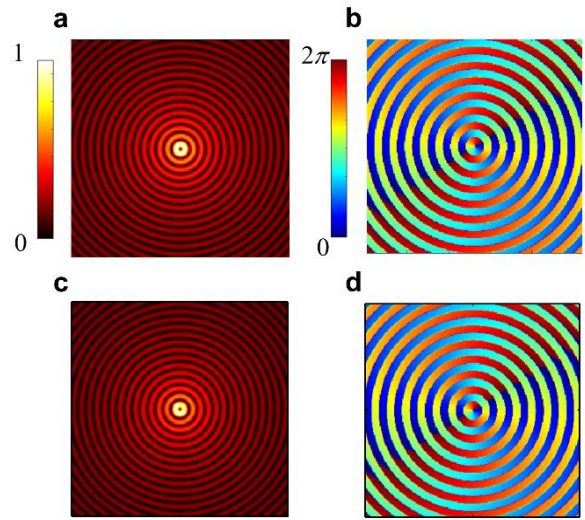

**Extended Data Figure 3 | Theory of photon-induced amplitude and phase modulation. (a),(b)** Calculated amplitude and phase of the out-of-plane electric field of a 1st order plasmonic Bessel vortex, created by a circular coupling slit as in fig. 2a. The field is calculated via the Huygens principle method[23]. **(c),(d)** Calculated amplitude and phase of the transverse electron wavefunction after interaction with the plasmonic vortex presented in **(a),(b)**, in the linear interaction regime. The wavefunction distribution is calculated via the expression given in the Methods section, by summing over the first 10 interaction orders. The fine match between the electron and electric field distributions suggests that light shapes both the electron amplitude and phase. As a specific consequence, angular momentum can indeed be transferred from the SPP field to the electrons interacting with it.

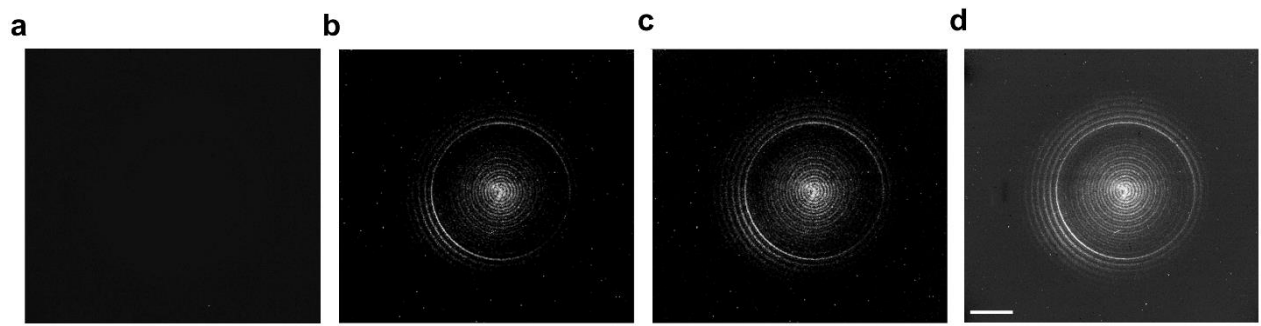

**Extended Data Figure 4 | Image processing of the electron distribution measurements.** The figure illustrates the process of creating the electron distribution images presented throughout the manuscript. **(a)** The raw data without any manipulation. Random pixel flaring greatly reduces image contrast. **(b)** Mitigation of random pixel flaring by contrast manipulation, as described in the Methods section. **(c)** Equalization of the image after contrast manipulation enables the visualization of more detailed features. **(d)** The image generated automatically from the detector software, similar in features to the images we present. The white scale bar in **(d)** is relevant for all images and corresponds to 5 microns.

**Acknowledgments.** Samples were prepared at the Technion's Micro & Nano Fabrication Unit, with the help of Dr. Guy Ankonina and Dr. Larisa Popilevsky. Measurements were conducted in I.K.'s UTEM laboratory in the electron microscopy center (MIKA) in the Department of Materials Science and Engineering of the Technion. The authors acknowledge the Russell Berrie Nanotechnology Institute and the Hellen Diller Quantum Center for their support of this research. S.T. acknowledges support by the Adams Fellowship Program of the Israel Academy of Science and Humanities, and wishes to thank A. Karnieli, A. Arie and Y. Kauffman for helpful conversations. The authors are especially grateful to the Q-SORT consortium for inspiring this work. I.K. acknowledges the support of the Azrieli Faculty Fellowship. This research was supported by the ERC starting grant NanoEP 851780, Research & Innovation grant SMART-electron 964591 and the Israel Science Foundation grants 3334/19 and 831/19.

**Data availability.** The data supporting the findings of this study are available from the corresponding author upon reasonable request.

**Author contributions.** S.T. and I.K. conceived the project. S.T. designed the experimental samples and performed their fabrication. R.D., S.T. and K.W. conducted the measurements. O.R. and S.T. performed simulations and theoretical calculations. G.B. and I.K. supervised the project. All authors participated in writing the manuscript and analyzing the experimental results.

**Competing interests.** The authors declare no competing interests.